\documentclass{article}
\usepackage[utf8]{inputenc}
\usepackage{INTERSPEECH2019}
\usepackage{amsmath,graphicx,url,times,booktabs, tabularx, multirow, adjustbox, dblfloatfix, comment, cite}
\usepackage{hyperref}

\title{Replay attack detection with complementary high-resolution information using end-to-end DNN for the ASVspoof 2019 Challenge}
\name{Jee-weon Jung$^{*}$\thanks{$^*$These authors contributed equally.},
    Hye-jin Shim$^{*}$,
    Hee-Soo Heo,
    and Ha-Jin Yu$^\dag$\thanks{$^\dag$ Corresponding author}\thanks{This research was supported by Basic Science Research Program
through the National Research Foundation of Korea(NRF) funded by the
Ministry of Science, ICT \& Future Planning(2017R1A2B4011609)}}

\address{School of Computer Science, University of Seoul, South Korea}
\email{jeewon.leo.jung@gmail.com,
    shimhz6.6@gmail.com,
    zhasgone@naver.com,
    hjyu@uos.ac.kr}

\begin{document}

\maketitle
\begin{abstract}
In this study, we concentrate on replacing the process of extracting hand-crafted acoustic feature with end-to-end DNN using complementary high-resolution spectrograms.
As a result of advance in audio devices, typical characteristics of a replayed speech based on conventional knowledge alter or diminish in unknown replay configurations.
Thus, it has become increasingly difficult to detect spoofed speech with a conventional knowledge-based approach. 
To detect unrevealed characteristics that reside in a replayed speech, we directly input spectrograms into an end-to-end DNN without knowledge-based intervention.
Explorations dealt in this study that differentiates from existing spectrogram-based systems are twofold: complementary information and high-resolution. 
Spectrograms with different information are explored, and it is shown that additional information such as the phase information can be complementary.
High-resolution spectrograms are employed with the assumption that the difference between a bona-fide and a replayed speech exists in the details. 
Additionally, to verify whether other features are complementary to spectrograms, we also examine raw waveform and an i-vector based system. 
Experiments conducted on the ASVspoof 2019 physical access challenge show promising results, where t-DCF and equal error rates are 0.0570 and 2.45 \% for the evaluation set, respectively.
\end{abstract}

\section{Introduction}
Automatic speaker verification (ASV) systems are being widely applied to various industries. 
However, spoofing attacks are becoming a threat to the reliability of ASV systems, necessitating the study of spoofing detection systems. 
Following this trend, the Automatic Speaker Verification Spoofing and Countermeasures (ASVspoof) initiative is providing a platform for researches to follow-up, study, and compare spoofing detection systems. 
The ASVspoof Challenge has covered various kinds of spoofing attacks, such as text-to-speech (TTS) and voice conversion (VC) in 2015, and replay attacks in 2017 \cite{wu2015asvspoof, kinnunen2017asvspoof}. 
ASVspoof2019 challenge deals with advances in TTS and VC technology as logical access and controlled simulation of replay attack as physical access \cite{todisco2019asvspoof}
VC and TTS require expertise and specialized equipment. 
In contrast, replay attacks does not require any expertise nor specialized equipment. 
It can be simply conducted by acquiring target speaker's voice using a recording device, and then replaying using a playback device. 
In this process, a different combination of replay and playback device with background environment can be used which is referred to as `replay configuration'. 
Despite the simplicity of attack scheme, replay attack has been proved as an effective way to deceive an ASV system.
This study concentrates on replay detection task. 

Through a survey on previous studies in replay detection including past ASVspoof competitions, we found that a number of researches have focused on finding discriminative features to improve spoofing detection \cite{font2017experimental, suthokumar2017independent, tom2018end, xiao2015spoofing}. 
Such features include constant Q cepstral coefficients (CQCC), inverse Mel-filter cepstral coefficients (IMFCC), linear prediction cepstral coefficients (LPCC), and group delay (GD)-grams. 
These features concentrate on representing the characteristics of a speech, that is considered discriminative in conventional knowledge for replay detection. 
For instance, IMFCC concentrates on high frequency bands, utilizing the knowledge that high frequency bands in replayed speech are often distorted.
However, as a result of advances in both recording and playback devices, distortion that reside in a replayed speech diminishes. 
We hypothesize that because of this phenomenon, discriminative power of conventional features will decrease. 

To deal with decreasing distortions in replayed speech, we explore an approach of minimizing the intervention of conventional knowledge and fully exploit DNN-based data driven approach. 
Our main focus in this study is to provide appropriate unprocessed, complementary information with high-resolution to facilitate end-to-end DNN. 
Complementary information combining not only general spectrograms which include magnitude information, but also phase information and power spectral density (PSD) is explored. 
We explore phase information, which has been shown to be effective in replay attack detection \cite{tom2018end, srinivas2018relative, li2018multiple, gunendradasan2018detection}, with PSD for concentrating on the distribution of the power signal over frequencies rather than concentrating on spectral contents.
To verify the effectiveness, we investigate model-level and score-level ensembles of various spectrograms with PSD. 
Experiments confirm that using complementary features contributes in the direct modeling of spectrogram-based deep neural networks (DNNs). 

Furthermore, we used high-resolution of 2048 fast Fourier transform (FFT) bins for all features. 
The purpose is being able to represent the subtle difference between bona-fide speech and spoofed speech. 
Because of the advancement of replay attacks, the difference may be more subtle, and less obvious, requiring focus on minute distinctions.
Our comparative experiments show that the resolution significantly affects actual performance (see Table 3). 

%The rest of this paper is organized as follows. 
%Section 2 describes the end-to-end DNN used in this study. 
%Section 3 describes the key components explored: complementary features and high-resolution. 
%Experiments and results are presented throughout Section 4 and 5, and the paper is concluded in Section 6. 

\begin{figure*}[ht!]
  \centering
  \includegraphics[width=0.8\textwidth]{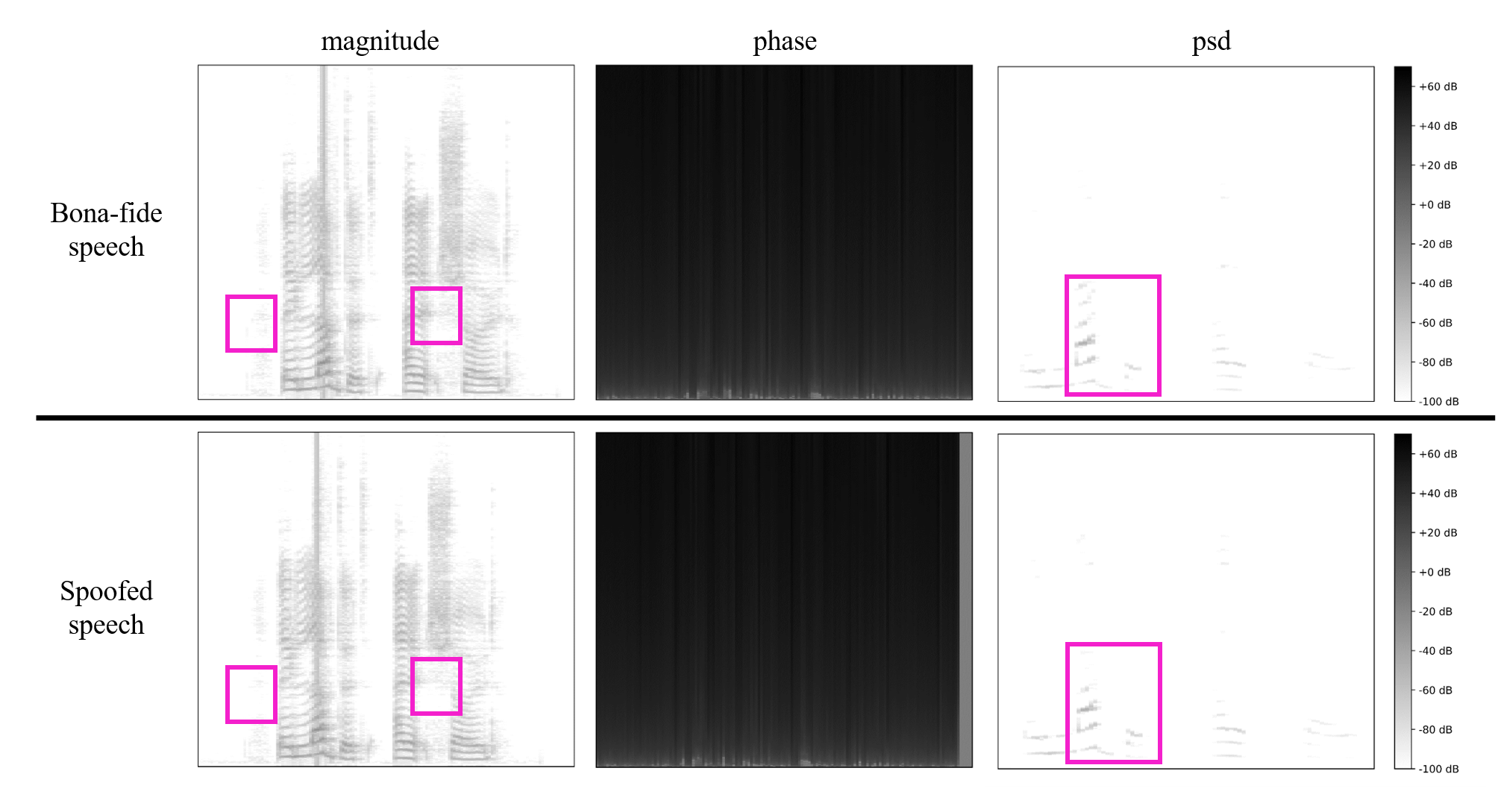}
  \caption{Visualization of bona-fide (upper) and replayed (lower) spectrograms and PSD: magnitude (left), phase (mid), and PSD (right). Minor differences in small regions (pink boxes) demonstrate the difficulty of replay attack spoofing detection task and the necessity of high-resolution.}
  \label{fig:1}
\end{figure*}

\section{End-to-end DNN}
We introduce an end-to-end DNN that is used to deal with decreasing distortions in replayed speech by minimizing the intervention of conventional knowledge. 
Using spectrograms as input, end-to-end DNN replaces the sub-process of selecting discriminative parts, which makes the intermediate representations adaptable to the data. 
An output of this model directly indicates a decision score when spectrograms are input, which simplifies the process pipeline. 
The state-of-the-art in various audio domain tasks has adopted an approach that utilizes a DNN that directly inputs spectrograms \cite{generalizedE2E, graves2014towards, chorowski2015attention, venugopalan2014translating}. 

The DNN used in this study comprises convolutional neural networks (CNNs), gated recurrent units (GRUs) and fully connected layers (CNN-GRU) as used in \cite{jung2018complete, jung2018avoiding, jung2018short}. 
In this architecture, input features are first processed using convolutional layers to extract frame-level embeddings. 
Convolutional layers comprise residual blocks \cite{Residual} with identity mapping \cite{full_pre} to facilitate the training of deep architectures. 
Specifically, the first convolutional layer of our model processes local adjacent time and frequency domains and is gradually aggregated by repeating pooling operations to extract frame-level embedding. 
Then, a GRU layer is employed to aggregate extracted frame-level features into a single utterance-level feature. 
One fully connected layer is used to transform the utterance-level feature. 
An output layer with two nodes indicates either the input utterance is bona-fide or spoofed.

\section{Complementary high-resolution feature}
In this section, we introduce the key aspects to facilitate training end-to-end DNN without intervention based on human priors; providing complementary information and using high-resolution. 
%However, because replay configurations such as the recording device, playback device, the environment, and the scenario can vary, these characteristics can be eliminated or decreased extent in unknown conditions with knowledge-based approach.
To generalize towards unknown replay configurations, we hypothesize that the approach of providing varied, raw information as input and performing data-driven feature selection with DNN would provide a more appropriate process of feature extraction for spoofing detection. 
Based on this hypothesis, spectrograms including various information with high-resolution are explored in expectation of outperforming acoustic features extracted with conventional knowledge. 
Specifically, phase information and PSD was exploited in addition to general spectrograms including magnitude information. 
In general, a spectrogram refers to a magnitude spectrogram containing absolute values of the fast Fourier transform (FFT). 
Whereas phase information was often overlooked in many audio domains, recent studies have demonstrated that phase-based features provide discriminative information for replay detection \cite{tom2018end, xiao2015spoofing, auger2012phase}. 
We use magnitude spectrogram and phase spectrogram supposing that phase information would supplement the magnitude information without requiring an additional extraction process, as both use partial information of FFT. 
We also exploit PSD for further improvement.
Because PSD concentrates on the distribution of signal power over frequencies, it differs from magnitude or phase, which concentrate on spectral contents.
Concurrently using PSD and spectrograms therefore enables consideration of the frequency distribution of the overall signal strength as well as the harmonics of amplitude and phase in the signal. 
To exploit this diverse information, we explore the combinations in both the model-level and score-level. 
Model-level ensemble inputs various features to a single DNN, whereas score-level ensemble exploits multiple DNNs and conducts score summation of DNNs' outputs, respectively.
To further analyze the relations between complementary information, we compared each combination using different spectrograms.
Figure 1 shows that, even with the same speech, different clues (pink boxes) for spoofing detection can be presented, depending on the types of spectrograms used.

As noted previously, with the advance in quality of audio devices, the difficulty of detecting spoofed utterances has been increased, as prominent differences between bona-fide speech and spoofed speech have been reduced. 
This necessitates the usage of high-resolution input that can be used to demonstrate the subtle differences residing in replay spoofed utterances. 
A high-resolution of 2,048 FFT bins was used for all spectrograms in this study. 
We were inspired by the experiments in \textit{Tom et al.} \cite{tom2018end} that attention-based GD-grams significantly outperformed spectrograms.
The GD-grams used in the \textit{Tom et al.} \cite{tom2018end} are obtained using 2,048 FFT bins, which had higher resolution than spectrograms.
We hypothesized that the difference in resolution could have also conducted a key role to the performance besides the difference of used feature. 
To verify this hypothesis, we conducted an comparative experiment. 
Results shown in Table 3 match our hypothesis where equal error rate (EER) of a spectrogram using 2,048 FFT bins significantly outperformed an identical system with 512 FFT bins.

\section{Experimental settings}
DNN training was implemented using Keras, a deep learning library for python, with a Tensorflow backend \cite{keras, tensorflow, tensorflow2}. 
The i-vector extraction was conducted using the Kaldi toolkit\cite{povey2011kaldi}. 

\subsection{Dataset}
We used the ASVspoof 2019 physical access dataset for all experiments. 
This dataset comprises 54,000 utterances as the training set, 29,700 utterances as the development set, and 137,457 utterances as the evaluation set. 
Utterances are recorded from 20 speakers (8 male, 12 female) at a 16-kHz sampling rate with 16-bit resolution. 
Training and development data comprises 27 different acoustic configurations using 3 room sizes, 3 levels of reverberation, and 3 speaker-to-ASV microphone distances.
9 different replay configurations are used, as combinations of 3 categories of attacker-to-talker recording distances and 3 categories of loudspeaker quality. 
The acoustic and replay configurations of the evaluation set are different from those of the training and development set. 

\begin{table}[t]
 \caption{DNN architecture ($l$: length of input sequence).}
  \centering
  \label{table:DNNarc}
  \begin{adjustbox}{width=0.8\columnwidth}
  \begin{tabular}{r c c c}
  \toprule
   layer & output shape  & kernel size  & stride \\
  \hline
  Conv1 & $l \times1024\times 16$ & $3\times 7$ &$1\times1$\\
  \hline
  Res1 & $ (l / 2) \times257\times 32$ & $3\times 5$ &$2\times4$\\
  \hline
  Res2 & $ (l / 4) \times65 \times 64$ & $3\times 5$ &$2\times4$\\
  \hline
  Res3 & $ (l / 8) \times17 \times 128$ & $3\times 5$ &$2\times4$\\
  \hline
  Pool & $ (l / 8) \times 1 \times 128$ & $1 \times 17$ & $ 1 \times 17$\\
  \hline
  GRU & $1 \times 512$ & - & -\\
  \hline
  Dense1 & $64$ & $512 \times 64$ & -\\
  \hline
  Output & $2$ & $ 64 \times 2 $ & -\\
  \bottomrule
  \end{tabular}
  \end{adjustbox}
\end{table}

\subsection{Spectrograms, raw waveforms, and i-vector extraction}
Spectrograms were extracted using a hamming window with a length of 50 ms and a shift size of 20 ms. 
Representation using magnitude, phase spectrogram, and PSD were extracted with 2,048 FFT bins each. 
The number of the time axis was fixed to 120 ($\approx$ 2.4 s), by either cropping long utterances or duplicating short utterances at the training phase for batch construction, depending on their lengths. 
Whole utterances were input at the evaluation phase without duration adjustment. 

Raw waveforms were directly input to the DNN without any pre-processing. 
The pre-emphasis layer was excluded, which differs from the setup in \cite{jung2018avoiding}, based on a comparison experiment. 
For batch construction, the length of each utterance was fixed to 26,244 samples ($\approx$ 1.64 s) by either performing random cropping for long utterances or duplicating for short utterances. 
At the evaluation phase, whole utterances were input to the DNN. 

The i-vectors were extracted using a universal background model with 256 diagonal Gaussian components which input 20-dimensional Mel-frequency cepstral coefficients with its first and second derivatives, comprising 60-dimensional acoustic features. 
200-dimensional i-vectors were extracted, and neither linear discriminant analysis nor length normalization were applied.

\subsection{DNN architecture}
A slightly modified ResNet was used for modeling the spectrograms, accounting for different stride sizes for time and frequency domains due to high-resolution in the frequency domain, and the number of residual blocks was adjusted to fit the provided ASV2019 physical access dataset. 
The raw waveform CNN-GRU model, proposed in \cite{jung2018short}, was used with a few modifications: one less residual block, a different specified input utterance length at training phase to fit the dataset, and additional loss functions for training (center loss \cite{wen2016discriminative} and speaker basis loss \cite{heo2019end}).                                       
This model first extracts 128-dimensional frame-level representations using 1-dimensional convolutional layers. 
Then a GRU layer with 512 nodes combines the extracted frame-level features into utterance-level features. 

A simple fully-connected DNN with 3 layers, each with 1,024 nodes, was used for i-vector modeling. 
For all DNNs, he normal initialization \cite{he2015delving}, weight decay with $\lambda=1e^{-4}$ was applied and trained with AMSGrad optimizer \cite{reddi2018convergence}. 
Additionally, for all systems, the output layer has two nodes, each indicating bona-fide and spoofed utterances. 
The output layer's node value that indicates a bona-fide utterance was directly used as the score (in end-to-end fashion) without additional modeling when an utterance was input.
The DNN architecture is summarized in Table \ref{table:DNNarc}\footnote{Implementation of raw waveform processing, spectrograms extraction, and DNN architecture are in \url{ https://github.com/\\Jungjee/ASV2019\_competition\_Jung}}\footnote{Modified model in PyTorch with training script is in \url{ https://github.com/Jungjee/ASVspoof2019\_PA}}.

\begin{table}[t]
  \renewcommand\thetable{2}
  \caption{Performance comparison between spectrogram-based systems with different types, raw waveform, and i-vector on the development set. Spectrogram-based with more than one type shows model-level ensemble results.}
  \label{tab:table3}
  \centering
  \begin{tabular}{lcc}
  \toprule
  \textbf{System} & \textbf{t-DCF}  & \textbf{EER (\%)}\\
  \midrule
  Baseline (CQCC-GMM) & 0.1953 & 9.87\\
  \hline
  Spec-magnitude & \textbf{0.0482} & \textbf{1.76}\\
  Spec-psd & 0.1153 & 3.74\\
  Spec-phase & 0.2145 & 8.04\\      
  Raw waveform   & 0.1915              & 8.03\\
  i-vector & 0.2119 & 8.74\\
  \toprule
  Spec-magnitude\&psd & \textbf{0.0491} & \textbf{1.75}\\
  Spec-magnitude\&phase & 0.0590 & 2.11\\
  Spec-psd\&phase & 0.1159 & 3.91\\
  Spec-magnitude\&psd\&phase & 0.0688 & 2.11\\
  \hline
  Spec-score-level ensemble & \textbf{0.0306} & \textbf{1.05}\\
  \bottomrule
  \end{tabular}
\end{table}
\begin{table*}[t] 
  \renewcommand\thetable{5}
  \caption{Performance comparison on the evaluation set using various attacker-to-talker distances and loudspeaker quality of the CQCC baseline and our submitted primary system. Two sets of labels refer to attacker-to-talker distance (A: 10-50 cm, B: 50-100 cm, C: far than 100 cm) and loudspeaker of quality (A: perfect, B: high, C: low) respectively.}
  \label{tab:tale1}
  \centering
  \begin{tabular}{lc cccc ccc ccc}
  \toprule
  Metric & System & \textbf{Pooled} & AA & AB & AC & BA & BB & BC & CA & CB & CC\\
  \midrule
  \multirow{2}{*}{t-DCF} & CQCC-baseline & 0.2454 & 0.4975 & 0.1751 & 0.0529 & 0.4658 & 0.1483 & 0.0433 & 0.5025 & 0.1360 & 0.0461\\
   & Primary & 0.0570 & 0.1603 & 0.0416 & 0.0207 & 0.0839 & 0.0232 & 0.0111 & 0.0529 & 0.0184 & 0.0081\\ 
  \midrule
  \multirow{2}{*}{EER}& CQCC-baseline & 11.04 & 25.28 & 6.16 & 2.13 & 21.87 & 5.26 & 1.61 & 21.10 & 4.70 & 1.79\\
  
   & Primary & 2.45 & 6.65 & 1.68 & 0.82 & 3.33 & 0.90 & 0.45 & 2.17 & 0.63 & 0.30\\
  \bottomrule
  \end{tabular}
\end{table*}
\begin{table}[t]
  \renewcommand\thetable{3}
\caption{Performance comparison of various FFT resolutions. Magnitude spectrogram, single best system, was used for comparison. In these experiments, window length and shift size were fixed to 30 ms and 10 ms respectively to ensure that the number of samples within a window is greater than nFFT. The performance difference of the nFFT 2,048 model with that of Table 3 is due to the different window length and shift size.}
\label{tab:table4}
\centering
\begin{tabular}{lccc}
\toprule
\textbf{System}  & \textbf{nFFT} & \textbf{t-DCF}  & \textbf{EER (\%)}\\
\midrule
\multirow{3}{*}{Spec-magnitude} & \textbf{2048} & \textbf{0.0894} & \textbf{3.07}\\
& 1024 & 0.1226 & 3.81\\
& 512 & 0.2488 & 7.83\\
\bottomrule
\end{tabular}
\end{table}

\section{Result analysis}
%Table 2 개별 시스템 결과 분석
In this section, we first evaluate the single systems, then verify the effect of using complementary features in model-level and score-level, and then demonstrate that high-resolution is necessary. 
First, the evaluations of single systems are shown in the $2^{nd}$ to $6^{th}$ rows of Table 2. 
All single systems clearly outperform the CQCC baseline. 
Magnitude spectrogram, which uses the absolute value of FFT, seems most appropriate for replay attack spoofing detection. 

%model-level ensemble은 안됐지만, score-level ensemble은 significant 성능 향상 확인 가능
Second, the effect of using complementary spectrograms is analyzed. 
The results of ensemble systems in model-level and score-level fusion are shown in the $7^{th}$ to $10^{th}$ rows and the $11^{th}$ row of Table 2, respectively. 
Model-level did not show improvement, but score-level resulted in significant improvement. 
Surprisingly, including model-level ensemble systems in score-level ensemble additionally brought further performance improvement where score-level ensemble of 7 spectrogram-based systems demonstrated an EER of 1.05 \%. 
To verify if other features can also complement various high-resolution spectrograms, we explored two more features. 
We explored raw waveform and i-vector because raw waveform does not include any pre-processing, and i-vector is a well-known utterance-level representation extracted based on human knowledge. 

%fft bin 개수에 따른 성능 비교
Next, Table 3 demonstrates the necessity of high-resolution by comparing performance with different numbers of FFT bins. 
Results show that high-resolution features are indeed critical for replay attack detection. 
Additionally, by comparing magnitude spectrogram systems with 2,048 FFT bins in Table 2 and Table 3, there was a considerable performance difference between spectrograms with a 50-ms window and a 20-ms shift in comparison to a 30-ms window and a 10 ms shift, where EERs were 1.76 \% and 3.07 \% respectively. 
Through this result, we note that the window length and shift size are also crucial for replay attack detection, as reported in \cite{muckenhirn2017long}.

%투고 시스템 소개
Table 4 shows the results of submitted systems for the ASV2019 physical access challenge. 
The magnitude spectrogram based system was submitted as Single. 
Based on the improvement brought with score-level ensemble, the primary system comprises spectrogram, raw waveform, and i-vector based models. 
The combined i-vector and 7 spectrograms based model was submitted as Contrastive1, and the combined raw waveform and 7 spectrograms based model was submitted as Contrastive2. 
Adding both raw waveform and i-vector further reduced EER to 0.96 \%, and was submitted as the Primary system for the competition. 

%세부 replay config에 대한 분석
Performance analysis of the baseline CQCC system and the `Primary' submission on different replay configurations, mainly for attacker-to-talker distance and replay device quality, is addressed in Table 5. 
Results demonstrate that the proposed system clearly outperforms the baseline regardless of replay configurations in terms of both t-DCF and EER. 
Although both attacker-to-talker distance and replay device quality affected the performance significantly, our `Primary' system was more robust towards replay spoofing using high quality devices. 
For replay attacks using high quality device (compare AA, BA, and CA), the baseline system consistently exhibited EER higher than 20 \% where our `Primary' submission could show improved performance as attacker-to-talker distance decreased. 
We interpret that using high-resolution played a key role for this result. 
\begin{table}[t]
  \renewcommand\thetable{4}
  \caption{t-DCF and EERs for the submissions to the ASV2019 physical access challenge condition development and evaluation set.}
  \label{tab:table5}
  \centering
  \begin{tabular}{lcccc}
  \toprule
  \multirow{2}{*}{\textbf{Submission}}            & \multicolumn{2}{c}{\textbf{t-DCF}}  & \multicolumn{2}{c}{\textbf{EER (\%)}}\\
  & \textbf{val}& \textbf{eval} & \textbf{val} & \textbf{eval}\\
  \midrule
  Primary     & \multirow{2}{*}{\textbf{0.0244}}& \multirow{2}{*}{\textbf{0.0570}} & \multirow{2}{*}{\textbf{0.96}} & \multirow{2}{*}{\textbf{2.45}}\\
  (7 spec+wave+i-vec)\\
  \hline
  Single & \multirow{2}{*}{0.0482} & \multirow{2}{*}{0.1255} &\multirow{2}{*}{1.76} & \multirow{2}{*}{4.79}\\
  (spec-mag)\\
  \hline
  Contrastive1 & \multirow{2}{*}{0.0246} & \multirow{2}{*}{0.0692} & \multirow{2}{*}{0.98} & \multirow{2}{*}{2.81}\\
  (7 spec+i-vec)\\
  \hline
  Contrastive2 & \multirow{2}{*}{0.0284} & \multirow{2}{*}{0.0632} & \multirow{2}{*}{1.10} & \multirow{2}{*}{2.73}\\
  (7spec+wave)\\
  \bottomrule
  \end{tabular}
\end{table}

\section{Conclusion}
In this study, we focus on replacing the hand-crafted feature extraction process by directly modeling spectrograms using DNNs in end-to-end fashion. 
As advanced recording and playback devices arise, characteristics of a speech, considered discriminative in conventional knowledge for replay detection diminish. 
Thus, it has become increasingly difficult to distinguish bona-fide speech from spoofed speech. 
To detect unrevealed characteristics that reside in a replayed speech, we directly input spectrograms into an end-to-end DNN without knowledge-based intervention. 
Utilizing explorations of this study such as complementary information and high-resolution further facilitates a data-driven approach. 
Additionally, the ensemble use of different features, including raw waveform and i-vector, was verified to further increase performance. 
The primary system submitted to the ASV2019 challenge demonstrated a t-DCF of 0.0570 and an EER of 2.45 \% and was compared to a t-DCF of 0.2454 and an EER of 11.04 \% baseline CQCC-GMM on the ASV2019 physical access challenge evaluation set. 

\newpage\newpage
\bibliographystyle{IEEEtran}
\bibliography{mybib}
\end{document}